\documentclass{article}
\usepackage{graphicx} 
\usepackage{bm,amsmath,color,dsfont, amssymb}
\usepackage[titletoc,toc,title]{appendix}
\usepackage{graphicx}
\usepackage[section]{placeins}
\usepackage{esint}
\usepackage[hidelinks]{hyperref}
\usepackage[justification=centering]{caption}
\usepackage{cleveref}
\usepackage{subcaption}
\usepackage{float}
\usepackage{multicol}
\usepackage{array} 
\usepackage{tabularx}
\usepackage{tgtermes} 
\usepackage{relsize}
\usepackage{wrapfig}
\usepackage{placeins}
\usepackage{authblk}

\usepackage{siunitx}
\usepackage{enumitem}
\usepackage{bm}
\usepackage{xcolor}
\usepackage{algorithm}
\usepackage{algpseudocode}
\usepackage{amsmath, amssymb}
\usepackage{mathtools}
\usepackage[
backend=biber,
style=nature,
]{biblatex}
\usepackage{booktabs}
\usepackage{multirow}

\newcommand{\eps}{\varepsilon}
\newcommand\red[1]{\textcolor{red}{#1}}

\DeclareMathOperator{\EX}{\mathbb{E}}

\title{The Subtle Interplay between Square-root Impact, \\ Order Imbalance  \& Volatility II: \\ An Artificial Market Generator}

\begin{document}

\author[1, 2, 3]{Guillaume Maitrier}
\author[3]{Grégoire Loeper}
\author[4, 2, 5]{Jean-Philippe Bouchaud}

\affil[1]{\textit{LadHyX UMR CNRS 7646, École polytechnique, 91128 Palaiseau, France}}
\affil[2]{\textit{Chair of Econophysics and Complex Systems, École polytechnique, 91128 Palaiseau, France}}
\affil[3]{
\textit{BNP Paribas Global Markets, 20 Boulevard des Italiens, 75009 Paris, France}}
\affil[4]{\textit{Capital Fund Management, 23-25 Rue de l'Université, 75007 Paris, France}
}
\affil[5]{\textit{Académie des Sciences, Paris 75006, France}}

\maketitle

\begin{abstract}
This work extends and complements our previous theoretical paper \cite{maitrier2025subtle} on the subtle interplay between impact, order flow and volatility. In the present paper, we generate synthetic market data following the specification of that paper and show that the approximations made there are actually justified, which provides quantitative support our conclusion that price volatility can be fully explained by the superposition of correlated metaorders which all impact prices, on average, as a square-root of executed volume. One of the most striking predictions of our model is the structure of the correlation between generalized order flow and returns, which is observed empirically and reproduced using our synthetic market generator. Furthermore, we were able to construct proxy metaorders from our simulated order flow that reproduce the square-root law of market impact, lending further credence to the proposal made in Ref. \cite{maitrier2025generatingrealisticmetaorderspublic} to measure the impact of real metaorders from tape data (i.e. anonymized trades), which was long thought to be impossible.  
\end{abstract}

\tableofcontents

\section{Introduction}

Market microstructure — and more specifically, limit order books — constitutes the microscopic environment in which prices are formed. It can be viewed as a black box: orders, submitted by various market participants, enter as inputs, and the resulting output is the observed transaction price. We describe it as a black box not because it is fundamentally opaque or inaccessible, but because the interactions that occur within it are governed by the actions of a multitude of heterogeneous agents, operating at different timescales, with diverse objectives and information sets. These interactions generate a highly nonlinear and noisy environment, making it extremely challenging to disentangle cause and effect, or to isolate the fundamental mechanisms driving price formation. For these reasons, understanding the inner workings of this black box — i.e. constructing models that faithfully reproduce both order flow patterns and price behavior — remains one of the central challenges in market microstructure research. Indeed, several known stylized facts about price impact (the famous ``square-root law'' ), order flow (with its long-memory properties) and volatility (i.e., that prices are diffusive) appear to be disconnected and, at least at first sight, hard to accommodate (for an in depth discussion, see \cite{bouchaud2018trades}). 

In our previous paper \cite{maitrier2025subtle}, we introduced a theoretical framework that aims to reconcile the statistical properties of order flow and price dynamics. Our model makes detailed and somewhat non-trivial predictions about the cross-correlations between order flow and price variations that appear to all be borne out by empirical data on stocks and futures.  


However, in order to derive such predictions, we made several assumptions and simplifications that may appear somewhat strong and uncontrolled \cite{maitrier2025subtle}. Whereas our theoretical model is challenging to solve analytically in full generality, it has the notable advantage of being rather straightforward to simulate numerically. The present follow-up paper serves a dual purpose. First, it offers additional evidence for the robustness of our theoretical model by showing that the approximate analytical treatment proposed in \cite{maitrier2025subtle} correctly describes the key empirical phenomena. Second, we introduce what we believe to be a versatile and realistic simulation tool that captures the intricate interplay between order flow and price formation, at least at the ``meso'' scale. 

The latter contribution could be of significant interest to the industry. Generating realistic market dynamics — encompassing both prices and order flow — remains a notoriously challenging task. It is fair to say that many existing approaches, including those based on neural networks, see \cite{theseSalma, coletta2022learning}, often fail to capture the full complexity of market behavior. However, such generative models are essential for several practical applications: they enable robust strategy back-testing, and they provide enhanced fitting capabilities in situations where real financial data is limited or unavailable. Our framework is based on a direct, mechanistic description of order flow and price impact that abstracts away from the infinite complexities of the full order book dynamics, and surely suffers from some short-comings, but is transparent and computationally trivial. Hybridizing our model with higher frequency, data driven generative model would be very interesting.  

This paper is divided in three parts, and contains: 
\begin{itemize}
    \item A detailed framework for generating synthetic data based on our assumptions. This synthetic dataset closely resembles the ideal one (similar to the TSE dataset, for instance \cite{maitrier2025double}) and includes all relevant information about order flow, metaorder IDs, execution time, and impact. 
    \item The reproduction of all empirical results studied in Ref. \cite{maitrier2025subtle}, but for now for simulated price, using parameters fitted on real data. 
    \item A discussion of the puzzling possibility of reconstructing metaorder proxies from public data \cite{maitrier2025generatingrealisticmetaorderspublic} that we confirm within our artificial market, thereby validating the procedure proposed in \cite{maitrier2025generatingrealisticmetaorderspublic} to measure the impact of metaorders without trader IDs. 
\end{itemize}

\section{A brief reminder of the generalized propagator model}\label{sec:recap_theory}

The present study is based on the unified framework proposed in \cite{maitrier2025subtle}. To succinctly recall the context, we summarize the model as follows:
\begin{itemize}
    \item The order flow is composed of a succession of metaorders, initiated with rate $\nu$ per unit time. The size (i.e., the number of child orders per metaorder) is distributed according to a power law, $\Psi_q(s)$, with a $q$-dependent tail exponent $\mu_q$, where $q$ is the size of the child orders, assumed to be constant within each metaorder. Such child volumes are distributed according to a lognormal distribution with parameters $(m, \sigma_\ell)$. To account for the empirical sign autocorrelations (see Fig. 3 of \cite{maitrier2025subtle} and Fig. \ref{fig:autocorr_child_CDV} below), we set $\mu_q = \mu_1 + \lambda \log(q)$.
    \item We also introduced the possibility of correlating the sign of {\it different} metaorders, starting respectively at time $t$ and $t+\tau$. If $\varepsilon_t$ is the sign of the $t^\text{th}$ metaorder of the day, we assume that for $\tau \gg 1$
    \begin{equation} \label{def_Gamma}
    \mathbb{E}[\varepsilon_t \varepsilon_{t+\tau}] = \Gamma \tau^{-\gamma_\times}.
    \end{equation}
    \item Finally, to understand price formation from order flow, we introduced a generalized propagator model. This instrument is crafted to incorporate the three main stylized characteristics of the impact of metaorders (see \cite{bouchaud2018trades, maitrier2025double} and refs therein): (i) impact grows on average as the square-root of the number of child orders being executed, (ii) average peak impact at the end of the execution solely depends on the square root of the traded volume, and (iii) average impact subsequently decays as a slow power-law of time after the end of execution. 
    
    We posited that the impact of a child order of volume $q$, executed at time $t'$ on the price at time $t > t'$, knowing that the metaorder started at $t=0$, is given by 
\begin{equation}\label{eq:new_propagator}
    G_q(t' \to t) = \frac{\theta \sqrt{q}}{(\varphi t' + n_0)^{1/2-\beta_q}} \left(\frac{\tau_0}{t-t' +\tau_0}\right)^{\beta_q}, \qquad (\beta_q < \frac12)
\end{equation}
with $\theta,n_0,\tau_0$ are constants — see section \ref{sec:how_to_simulate} for details —  and $\varphi$ the participation rate of the metaorder. Empirical observations led us to the following specification $\beta_q := \beta_1-\lambda'\log(q)$, meaning that impact decay is slower for large child orders, as intuitively meaningful.  
\end{itemize}

The entire framework is motivated and explained more thoroughly in \cite{maitrier2025subtle}, and leads to the following predictions: 

\paragraph{1. The generalized order flow imbalance:} We defined the weighted order flow imbalance, where $\eps_t$ is the sign of \textit{child orders}:
\begin{equation}\label{eq:Ia}
    I^a_T = \int_0^T \mathrm{d}t\,\eps_t q_t^a,
\end{equation}
and its moments $\Sigma_{I^a}^{(2n)}:= \mathbb{E}[(I^a_T)^{2n}]$, for which our theory predicts a non-trivial behavior: 
\begin{equation}\label{eq:scaling_Ia_n}
    \Sigma_{a,1}^{(2n)} \propto 
    \begin{cases} 
        T^{2n + 1 - \mu_m - 2na \lambda \sigma_\ell^2}, & \quad a <  a_c(n); \\
        T, & \quad a \geq a_c(n),
    \end{cases}
\end{equation}
with $a_c(n)=(1 - \mu_m/2n)/\lambda \sigma_\ell^2$.

\paragraph{2. The time-dependent covariance function:} Armed with the order flow description and the generalized propagator, we describe the interplay between price returns $\Delta_T$ and order flow by computing the covariance $\EX[\Delta_T\cdot I^a_T]$. Our model tells us that such a quantity should behave as a power-law of $T$ with an exponent that is a {\it non-monotonic} function of $a$:
\begin{equation} \label{eq:scaling_IDelta_q}
     \mathbb{E}_q[\Delta_T \cdot I_T^a] \propto  \begin{cases}
        & T^{5/2 - \widehat \mu(a)}, \qquad \widehat \mu(a) = \mu_m + (a + \frac12) \lambda \sigma_\ell^2 \qquad \qquad a < a_c' ;\\
        & T^{1 - \widehat \beta(a)} , \quad \qquad \widehat \beta(a) = \beta_m -  \left(a + \frac12\right)\lambda' \sigma_\ell^2  \qquad \quad a > a_c',
    \end{cases}
\end{equation}
where $\mu_m = \mu_1 + \lambda m$, $\beta_m=\beta_1 - \lambda' m$ and $a_c'$ such that $\widehat \mu(a_c') = \mu_{q_c'}$, with $5/2 - \mu_{q_c'} = 1 - \beta_{q_c'}$. 

\paragraph{3. The correlation coefficient:} Finally, our model also allows one to predict the behavior of the following correlation coefficient:

\begin{equation}\label{eq:corr}
    R_a(T) := \frac{\mathbb{E}[\Delta_T \cdot I_T^a]}{\Sigma_T \Sigma_{I^a}}, \qquad \Sigma_T:= \sqrt{\mathbb{E}[\Delta_T^2]}, \qquad \Sigma_{I^a}:=\sqrt{\mathbb{E}[(I_T^{a})^2]}
\end{equation}
The following prediction fits surprisingly well empirical data : 
\begin{equation} \label{eq:R_a_vs_a}
    R_a(T) =  e^{-\frac{\sigma_\ell^2 a^2}{2}} \left(A(T) e^{\frac{\sigma_\ell^2 a}{2}} + B(T)  e^{\lambda \sigma_\ell^2 a\log T }\right),
\end{equation}
for $a< a_c$, and $A,B$ two functions of $T$. In particular, for a given $T$, $R_a(T)$ is non-monotonic in $a$ and reaches a maximum for $a \approx 1/2$ for stocks and $a \approx 1$ for futures. 

Although the model is based on only a few assumptions, the theoretical predictions above are not straightforward, and some uncontrolled approximations needed to be made. Still, the empirical data we analyzed in \cite{maitrier2025subtle} agree surprisingly well with our predictions. By simulating numerically the very same model, our goal is to replicate these stylized facts without any analytical approximations, and demonstrate that we have identified the correct mechanism. This will confirm that such good fits are not merely coincidental and that uncontrolled approximations are not, unwittingly, responsible for the success of our theory.

\section{How to simulate our model?}\label{sec:how_to_simulate}

Whereas generating order imbalances is relatively straightforward, simulating realistic price dynamics is more delicate. In our model, child orders from different metaorders can in principle be executed simultaneously, which complicates the price formation process. Furthermore, while the execution of a child order is clearly a discrete event, its impact decays continuously over time and should be taken into account at each timestep.

After testing several approaches, we found that using actual timestamps yields the most transparent and realistic simulations. The simulation procedure is thus divided into three main steps: 
\begin{itemize}
    \item \textbf{Generating metaorders: } We specify the average number of metaorders and the total trading period for the simulation (e.g., 10000 metaorders over an 8-hour trading day). This defines the rate $\nu$ at which new metaorders start. For each metaorder, we define the following parameters: a volume $q$, distributed as a log-normal truncated below $q=1$, a size $s$ (distributed according to $\Psi_q$), a sign (either randomly assigned or generated with cross-metaorder correlations), and a starting time, randomly chosen within the trading day with density $\nu$.  \textit{We ensure that starting times are unique, as they will later serve as metaorder identifiers}. It is possible to control the trading rate and liquidity by modifying $\nu,\varphi,m,\sigma_\ell$, as described in section \ref{sec:recap_theory}.

    \item \textbf{Deriving the corresponding order flow : }
    The order flow is then generated by iterating over all time-sorted metaorders. For each one, we store the execution  time of child orders by generating time intervals $\delta t$ thanks to a Poisson process: $\delta t \sim e^{-\varphi \delta t}$. For example, the second child order is executed at time $t = t_{\text{start}} + dt_1$.
    This approach allows us to sort all child orders by their execution time, thereby constructing an order flow that closely resembles real trade-by-trade data (or more precisely that from the TSE dataset). Each event (here execution) includes the timestamp, volume, sign, child order rank, and the time elapsed $\tau$ since the start of the corresponding metaorder. In addition, and specific to our model, we store the value of $\beta_q$ associated with each metaorder.

\begin{table}[H]
\centering
\begin{tabular}{@{}llllll@{}}
\toprule
\textbf{timestamp} & \textbf{volume} & \textbf{sign} & \textbf{rank} & $\textbf{timestamp}_{\text{start\_meta}}$ & $\beta_q$ \\ \midrule
10:32:01.35 & 10 & +1 & 1 & 10:32:01.35 & 0.31 \\
10:32:01.57 & 150 & -1 & 7 & 09:15:03.86& 0.20\\
10:32:02.15 & 80  & +1 & 2 & 09:34:43.12& 0.25\\
10:32:02.76 & 120 & -1 & 1 & 10:32:02.76& 0.22\\
10:32:02.78 & 90  & -1 & 5 & 09:47:52.27& 0.28\\ \bottomrule
\end{tabular}
\caption{Simulated order flow data with metaorder decomposition. Each row represents the execution of a child order, with \textit{'rank'} column indicating the position of the child order within its metaorder. The \textit{'timestamp(start\_meta)'} column records the start time of the metaorder and also acts as an identifier, as it is uniquely assigned to each metaorder}
\end{table}

    \item \textbf{Reconstructing the mid-price: }
    Armed with this simulated order flow and our generalized propagator, reconstructing the price dynamics becomes straightforward. We define the price $p_t$ as the mid-price \textit{just before} the execution occurring at time $t$. To compute this price, we aggregate the contributions from all child orders executed prior to $t$, ie $t_{\text{exec}}<t$. We use the generalized propagator to compute for their respective impacts and sum them. Although not computationally optimized (with complexity $\sim \mathcal{O}(N^2)$), this algorithm appears to be the most rigorous. It also preserves a key property of price impact observed in real markets: most of real market orders have zero immediate impact (as their volume is smaller than the prevailing best), but their impact builds up over time (on this point, see e.g. \cite{bouchaud2003fluctuations, bouchaud2018trades, maitrier2025double}).
\end{itemize}

To complement this description, we provide the following pseudo-code :

\begin{algorithm}[H]
\caption{Simulation of Price Impact from Correlated Metaorders}
\begin{algorithmic}[1]
\Require Number of metaorders $N$ and base parameters $\gamma_\times$, $\mu_1$, $\beta_1$, $(m, \sigma_\ell)$, ($\lambda$, $\lambda'$)
\Ensure Time series of executed orders with associated impact prices
\vspace{0.5em}

\State Set $\nu$, the Poisson rate for metaorders initiation and $\varphi$ the participation rate within a metaorder.
\State Draw $N$ metaorder start times $\{t_i^{\text{start}}\}_{i=1}^N$, with $t_{i+1}^\text{start}-t_i^{\text{start}} \sim \text{Exp}(-\nu dt)$ 
\For{$i = 1$ to $N$}
    \State Sample metaorder volume $q_i \sim LN(m, \sigma_\ell)$ and size $\mu_i = \mu(q_i, \mu_1, \lambda)$ 
    \State Compute impact exponent $\beta_i = \beta(q_i, \beta_1, \lambda')$
    \State Sample metaorder sign $\varepsilon_i$ autocorrelated sign time serie, if $\gamma_\times$
    \State Sample number of child orders $s_i \sim \Psi_{q_i}$
    \State Generate inter-arrival times $\{\delta t^{(i)}_k\}_{k=1}^{s_i - 1} \sim \text{Exp}(-\varphi \delta t)$
    \State Compute execution times $t^{(i)}_k = t_i^{\text{start}} + \sum_{j=1}^{k} \delta t^{(i)}_j$
    \State Store each child order as $(t^{(i)}_k, \varepsilon_i, q_i, t_i^{\text{start}}, \beta_i, \mu_i)$
\EndFor

\State Sort all child orders by execution time $\{t_k\}$
\State Initialize price impact array $p_k \gets 0$
\For{each execution time $t_k$}
    \State Identify past orders $j < k$
    \State Apply the generalized propagator:
    \[
    p_k = \sum_{j < k}
    \varepsilon_j \cdot \sqrt{q_j} \left(\varphi(t_j - t_j^{\text{start}}) + n_0\right)^{-\frac12 + \beta_j} \cdot
    \left( \frac{\tau_0}{t_k - t_j + \tau_0} \right)^{\beta_j}
    \]
\EndFor

\State Convert timestamps to realistic time
\State \Return DataFrame of child orders with $\{t_k, p_k, q_k, t_k^{\text{start}}, \varepsilon_k, \beta_k\}$
\end{algorithmic}
\end{algorithm}

This simple model relies on only a few parameters that require fine-tuning. To stay as close as possible to \cite{maitrier2025subtle}, we set $m \in \{3, 6\}$, $\sigma_\ell = 1$, $\lambda \sigma_\ell^2 = \frac{1}{8}$, and $\lambda' = 2\lambda$. We also set $\mu_m = 1.5$, and  $\beta_m = 0.25$. For consistency, we ensure that $0 < \beta_q < 1$. 

Finally, we fix $n_0 = 3$, based on empirical observations in \cite{maitrier2025double}, after verifying that this parameter has only a mild influence on the rest of the system. The average time between two child orders, denoted $\tau_0$, is theoretically given by $\tau_0 := (\nu \varphi \bar{s})^{-1}$ \cite{maitrier2025subtle}. For simplicity, we assume a uniform participation rate $\varphi$ across metaorders. By adjusting $\nu$ and $\varphi$, one can control the average number of concurrently active metaorders. In the rest of the paper, we will typically impose $\nu = 1.5 \cdot 10^{-3}$ and $\varphi=2\cdot 10^{-3}$.

\section{Empirical stylized facts vs. simulations}

We simulated the system under five different scenarios in order explore the relative importance of metaorder correlation, child volume fluctuations and the $q$-dependence of exponents $\beta$ and $\mu$.
We summarize the different names of these specifications in Table \ref{tab:simulation_cases}.

\begin{table}[H]
\centering
\begin{tabular}{|c|c|c|c|}
\hline
\textbf{Name} & \textbf{Metaorder Correlation} & \textbf{$q$-Dependence} & \textbf{$q$-Fluctuations }  \\
\hline
NC-NVD-NVF   & $\Gamma=0$  & $\lambda, \lambda'=0 $  & $q \equiv 1$ \\
NC-NVD-VF    & $\Gamma=0$  & $\lambda, \lambda'=0 $  & $LN(m,\sigma_\ell)$ \\
NC-VD-VF     & $\Gamma=0$  & $\lambda, \lambda' \neq 0 $ & $LN(m,\sigma_\ell)$ \\
C-NVD-VF     & $\Gamma > 0$ & $\lambda, \lambda'=0 $  & $LN(m,\sigma_\ell)$ \\
C-VD-VF      & $\Gamma> 0$ & $\lambda, \lambda' \neq 0 $ & $LN(m,\sigma_\ell)$ \\
\hline
\end{tabular}
\caption{Summary of the five simulated configurations. Each model is named using a triplet notation, with C = correlation (described by parameter $\Gamma$), VD = volume dependence of $\mu_q$, $\beta_q$, VF = volume fluctuations. Here, ''N" indicates negation, such as ND = no metaorder correlation ($\Gamma=0$, see Eq. \eqref{def_Gamma}), NVD = no volume dependence ($\lambda, \lambda'=0 $), NVF = no volume fluctuations ($\sigma_\ell=0$). The fully realistic case corresponds to the last line C-VD-VF. }
\label{tab:simulation_cases}
\end{table}

\subsection{The $q$-dependence of the autocorrelation of trades}

\begin{figure}[H]
    \centering
    \includegraphics[width=0.8\linewidth]{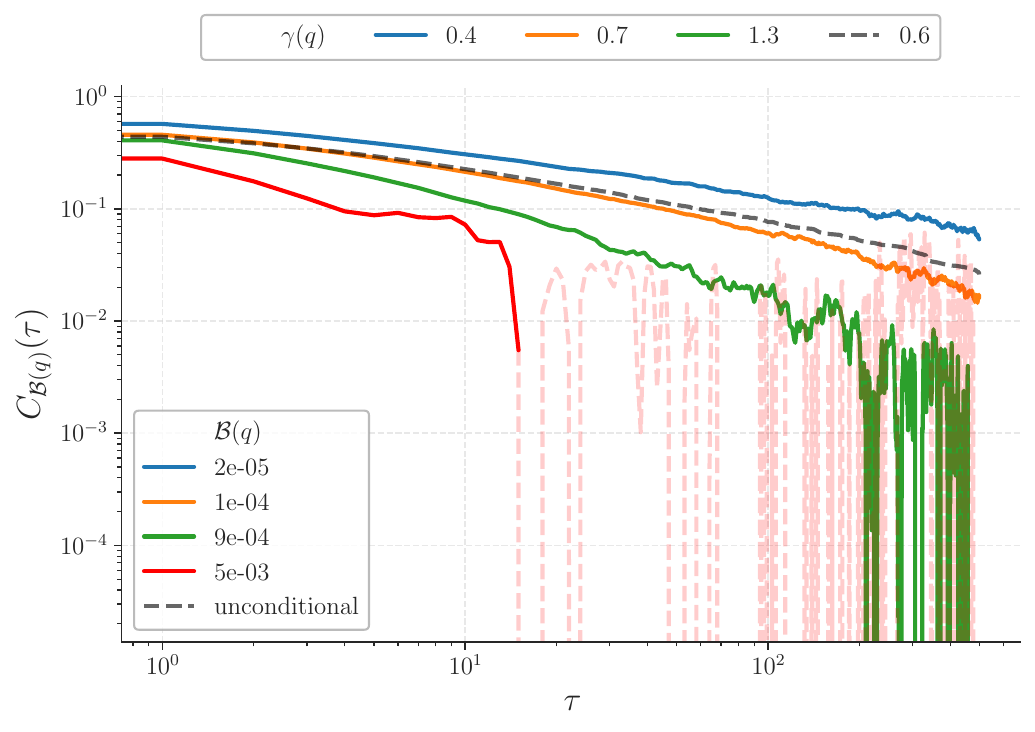}
    \caption{Evolution of the sign autocorrelation of market orders based on their corresponding volume bin $\mathcal{B}(q)$. Simulation were done in the C-VD-VF case, with $m=3,\sigma_\ell=1$ and $\lambda \sigma_\ell^2 = 1/8$. The dotted line corresponds to the {\it unconditional} autocorrelation function. Compare with Fig 3. in \cite{maitrier2025subtle}.}
    \label{fig:autocorr_child_CDV}
\end{figure}

We begin by examining the relationship between child order volume and their autocorrelation in the C-VD-VF scenario, which captures all effects we purport are important. To this end, we partition the simulated rescaled volume $\tilde q = q / \phi_D$, where $\phi_D$ denotes the daily traded volume, into four logarithmic bins $\mathcal{B}$. For each bin, we compute the sign autocorrelation function defined as 
\[
C_{\mathcal{B}(\tilde q)}(\tau) = \mathbb{E}[\varepsilon_{\mathcal{B}(\tilde q)}(t)\varepsilon_{\mathcal{B}(\tilde q)}(t+\tau)] \propto \tau^{-\gamma(q)}
\]
 The autocorrelation functions are displayed in Fig.~\ref{fig:autocorr_child_CDV}, in log-log scale, along with the unconditional autocorrelation function (dotted line). As observed in the data \cite{maitrier2025subtle}, the effective memory exponent $\gamma(q)$ systematically increases with volume, ranging from $0.4$ (long memory) to $1.3$ (short memory). This graph is strikingly similar to the one obtained for the EUROSTOXX, see Fig 3. in \cite{maitrier2025subtle}.

It is straightforward to verify numerically that the $q$-dependence of $\mu_q$ is indeed responsible for this phenomenon. If the order flow is simulated without incorporating this dependence, the stylized fact completely disappears, with $\gamma(q) \approx 0.5$ independently of $q$ (data not shown).

\subsection{The scaling of the order flow imbalance}

\begin{figure}[H]
    \centering
    \includegraphics[width=0.9\linewidth]{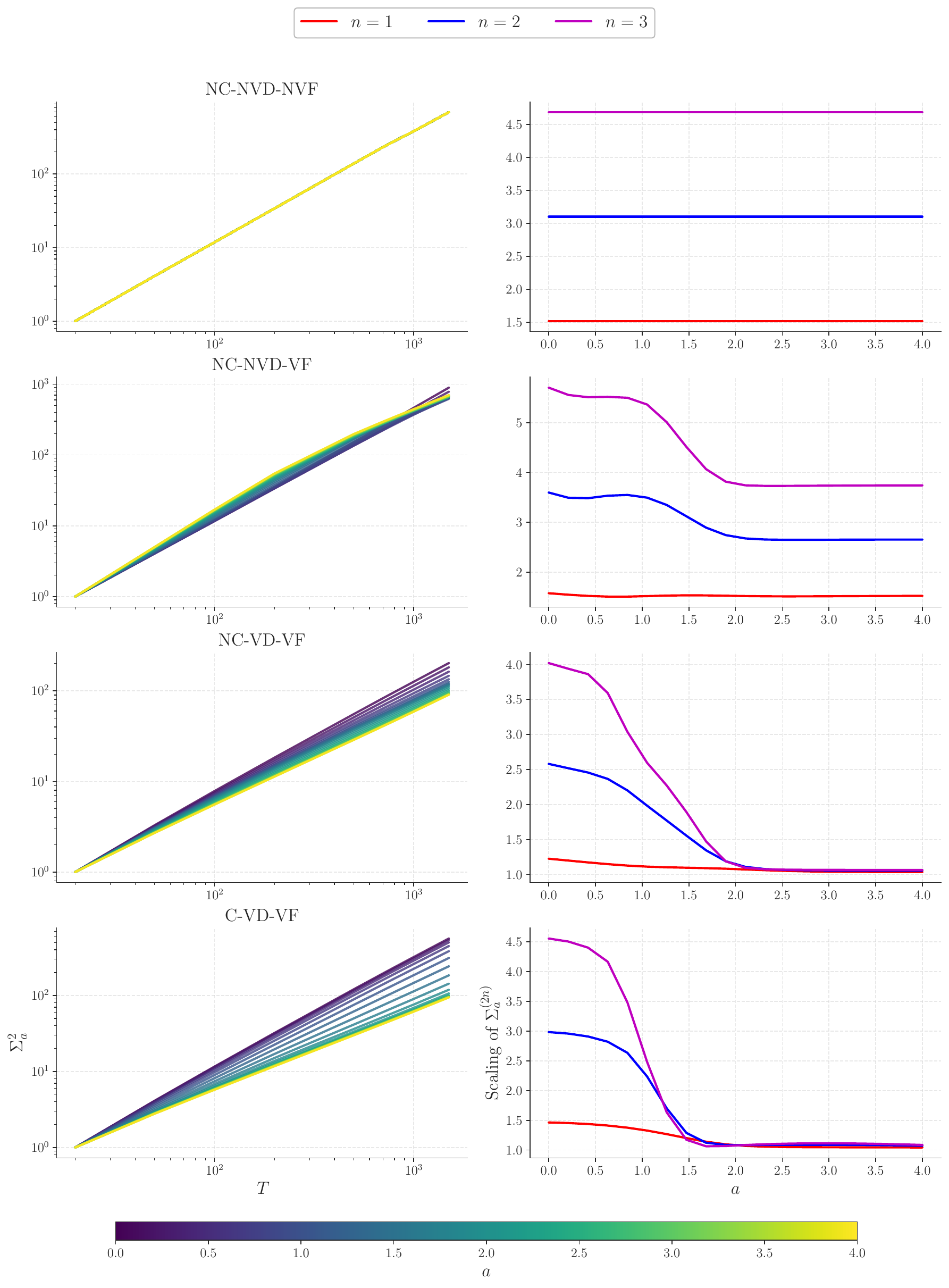}
   \caption{\textbf{Right column:} Scaling behavior of the moments $\Sigma_{a}^{(2n)}$ as a function of trade time $T$, from which the scaling exponent is extracted via a log-log regression. \textbf{Left column:} Scaling exponent plotted as a function of $a$. As predicted by our model, increasing $a$—which gives greater weight to large-volume orders—reduces the scaling exponent. We set $m=6, \sigma_\ell=1$ and $\lambda \sigma_\ell^2 = 1/8$.}
    \label{fig:simuIa_m6}
\end{figure}

We now turn to the scaling behavior of the moments of the generalized order imbalance $\Sigma_{I^a}^{(2n)}$, which is one of the main successes of the theoretical framework introduced in \cite{maitrier2025subtle}. When $q$ is constant, the dependence on $a$ disappears trivially, and the imbalance was shown in \cite{maitrier2025subtle} to follow a truncated Lévy distribution, entirely driven by the long memory of trade signs, thereby justifying the scaling $\Sigma_{I^a}^{2} \sim T^{3-\mu}$ with $\mu = 1.5$ for the NC-NVD-NVF simulation. However, by introducing a $q$-dependent $\mu_q$ (i.e. when $\lambda > 0$), we retrieve scalings that resemble very closely empirical ones, see Fig. \ref{fig:simuIa_m6}, both with (C) and without (NC) metaorder correlations, as expected.

Note that volume fluctuations alone can induce a spurious dependence of the scaling exponent on $a$ (see NC-NVD-VF in Fig. \ref{fig:simuIa_m6}) which is due to finite size effects, for which extreme events are artificially amplified as $a$ increases, with a mechanism similar to the Random Energy Model (REM) in spin glass theory \cite{derrida1981random, bouchaud1997universality}. Indeed, we only simulated 100 trading days with approximately 50000 trades each day such that these finite-size effects are noticeable.  

\subsection{Recovering a diffusive price}

A well known puzzle in the literature is the compatibility of decaying square-root impact, long-memory of trade signs and the diffusivity of prices — see \cite{bouchaud2003fluctuations, lillo2004long, taranto2018linear, bouchaud2018trades, sato2025exactlysolvablemodelsquareroot}. Several solutions to this conundrum were proposed in Section 4 of Ref. \cite{maitrier2025subtle}. In particular (i) the sign of metaorders themselves should be long-range correlated, as in Eq. \eqref{def_Gamma} and (ii) large child orders tend to have a permanent impact, i.e. beyond some value called $q_0$ in \cite{maitrier2025subtle}, the decay exponent $\beta_q$ becomes zero. 

These two scenarii are both confirmed by numerical simulations: we indeed find that the generalized propagator model leads to a sub-diffusive price in the absence of metaorder correlations ($\Gamma=0$) and without volume effects. Introducing metaorder autocorrelations with the correct exponent $\gamma_\times$ or incorporating a volume dependence $\beta_q$ restores price diffusivity at long times. By correctly tuning $\Gamma$ and $\lambda'$, one can control the full signature plot and not only the long time diffusive behaviour, and reproduce empirical results that show a variety of possibly short time behaviour, from locally trending to locally mean-reverting — although tick size effects, not modeled here, are expected to play an important role at short times.  

\begin{figure}[H]
    \centering
        \includegraphics[width=0.7\linewidth]{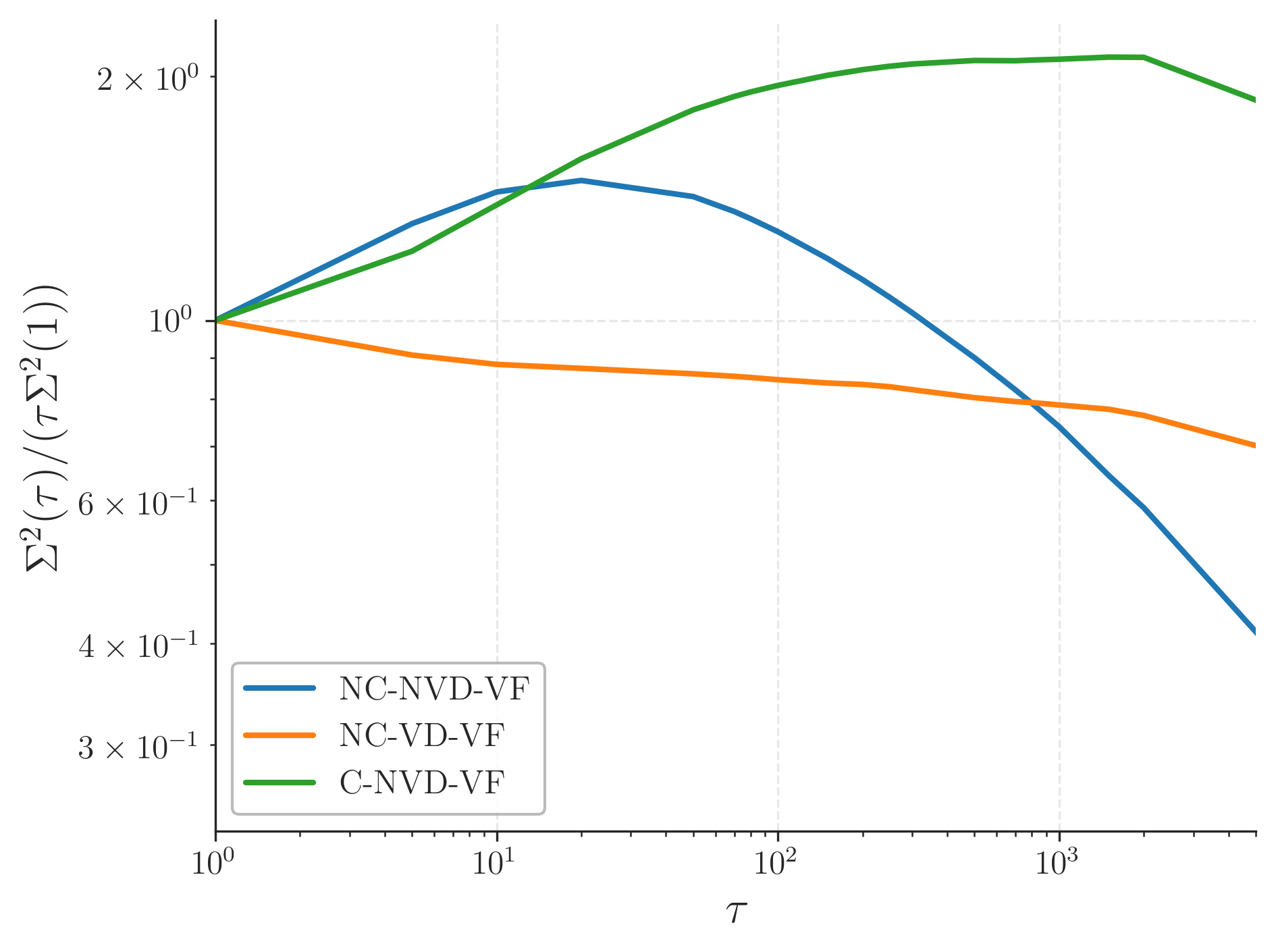}
    \caption{Signature plot $\Sigma^2/\tau$ of the simulated price  as a function of the trade lag $\tau$. Diffusion corresponds to a flat, horizontal signature plot. The generalized propagator model NC-NVD-VF (blue curve) results in sub diffusive behavior, as expected, while the two other impact models exhibit diffusive behavior after an initial trending phase (C-NVD-VF, green line) or mean-reverting phase (NC-VD-VF, orange line). Simulations were conducted for $\Gamma=0.1$ in the C-NVD-VF case, and $\lambda = \lambda'=1/6$ for the NC-VD-VF case. In both cases, we used $\varphi=2\cdot10^{-3}, \mu_m = 1.5, m = 3$ and  $\sigma_\ell^2=1$}
    \label{fig:diffusion_two_timescale}
\end{figure}

\begin{figure}[H]
    \centering
    \includegraphics[width=\linewidth]{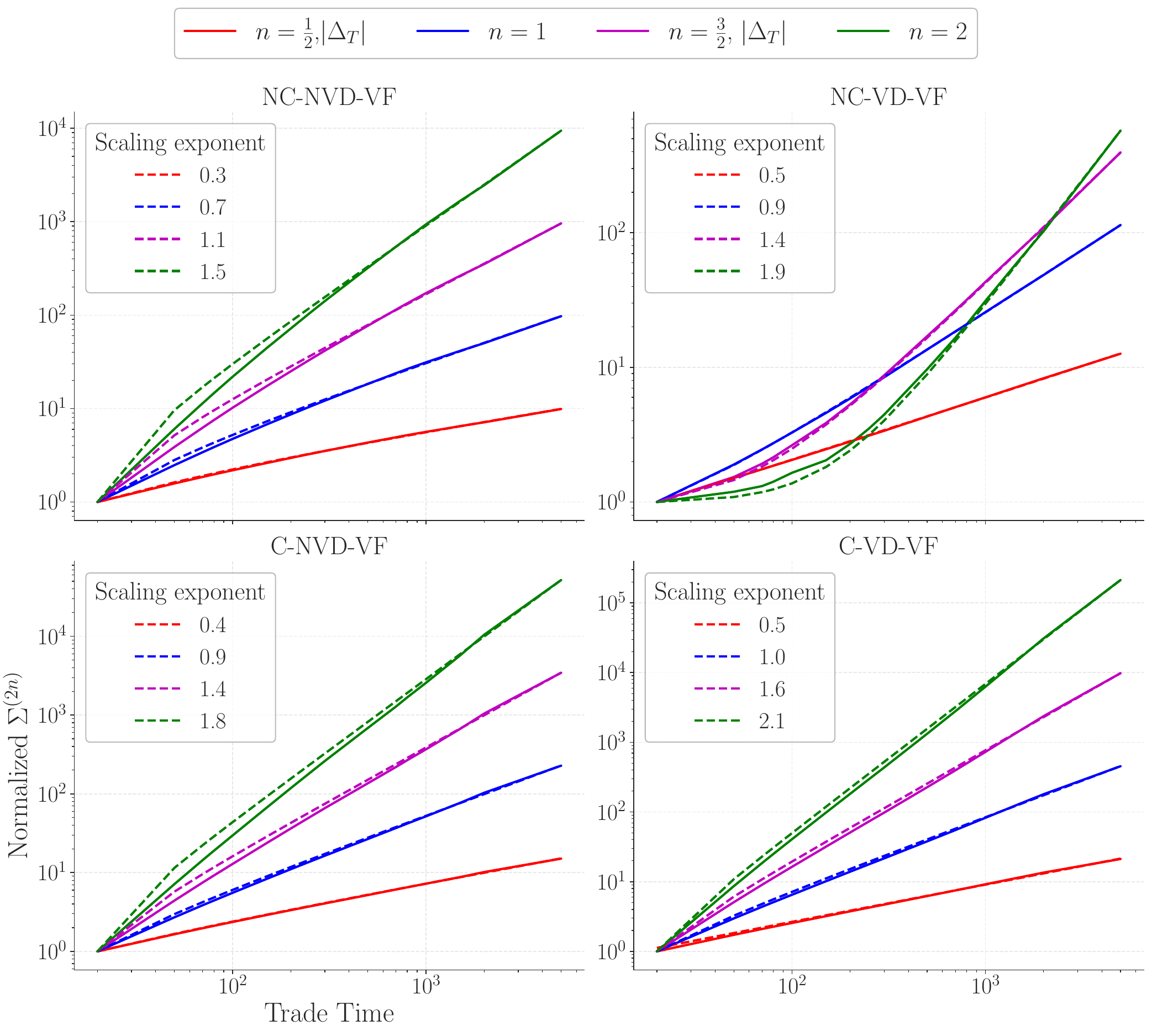}
    \caption{Scaling of the moments of price changes $|\Delta_T|^{2n}$ as a function of trade time $T$. We normalized the moment values such that all curves begin at $1$ for $T=1$. To account for short term, we fitted the data as $\Sigma^{(2n)} = a_0 + a_1 T^{\zeta_n} $ and present the values of $\zeta_n$ in the legend.}
    \label{fig:scaling moments}
\end{figure}

As in \cite{maitrier2025subtle}, we can also investigate 2n-moments of price changes, and check that all moments scale asymptotically as $T^{n}$, as for empirical data, see Fig. \ref{fig:scaling moments}. We insist again that we work here in trade time, so that multifractal effects due to intermittent activity bursts, are not present. 

\subsection{Aggregated impact and anomalous rescaling}

Aggregated impact is a very natural observable to investigate, but it also turns out to be highly non trivial. It is defined as the conditional expectation $\mathbb{E}[\Delta | I^a]$ of price change $\Delta$ given an imbalance $I^a$, is a natural and empirically accessible observable \cite{plerou2002quantifying, patzelt2018universal}. However, it exhibits non-trivial behavior that departs from the standard square-root law, with scaling properties that vary significantly with the time horizon $T$.

In particular, for $a=0$, the initial slope of $\mathbb{E}[\Delta | I^0]$ scales as $T^{-\omega}$ with $\omega \approx 1/4$, a result documented in \cite{patzelt2018universal, bouchaud2018trades}. While a Gaussian assumption would suggest a linear relation 
\begin{equation} \label{eq:cond_I}
    \mathbb{E}[\Delta | I^a] = \frac{\mathbb{E}[\Delta \cdot I^a]}{\Sigma_{I^a}^2} I^a,
\end{equation}
such an approximation has {\it a priori} no reason to hold within in our setting, where $I^a$ is a truncated Lévy variable. Despite this, Eq.~\eqref{eq:cond_I} still captures the correct $T$-scaling.

\begin{figure}[H]
    \centering
    \includegraphics[width=0.7\linewidth]{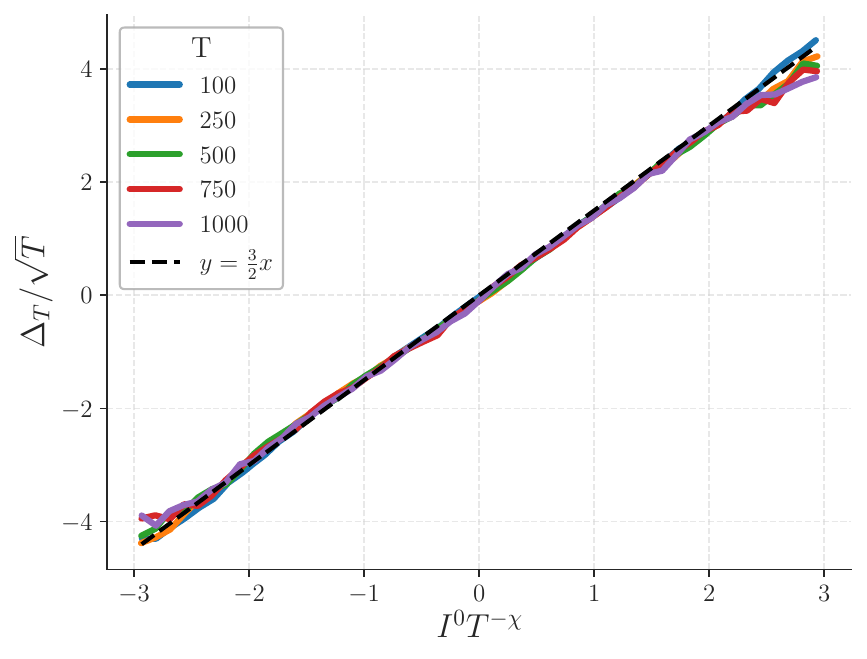}
    \caption{Aggregated impact as a function of sign imbalance for C-ND-V simulations. As in real market data, curves corresponding to different values of $T$ collapse onto a single master curve after appropriate rescaling. We find a scaling exponent $\chi = 0.75$, in close agreement with the theoretical prediction $1/\mu$, as we simulated with $\mu = 1.5$. The slope exponent $\omega = 0.25$ is also consistent with empirical observations.}
    \label{fig:Felix_simu}
\end{figure}

We now revisit this observable using simulations based on our model and confirm that the anomalous rescaling $\sim T^{-\omega}$ is precisely recovered, validating the theoretical prediction. However, Fig.~\ref{fig:Felix_simu} shows that the concavity seen in empirical curves for large imbalances is absent in our simulations. As demonstrated in Ref. \cite{patzelt2018universal}, such a concavity is due to a selection bias, not described in our model: large orders tend to be executed when large limit orders are available on the other side, limiting the impact of these market orders.

\subsection{The covariance coefficient}

We now focus on the covariance coefficient. Our theoretical predictions suggest that the non-monotonic shape as a function of $a$ originates from volume fluctuations — particularly in the upward branch, which depends on the parameter $ \lambda' $ in the relation $ \beta(q) = \beta_1 - \lambda' q $ (see Eq. \eqref{eq:scaling_IDelta_q}). We clearly confirm this phenomenon in Fig. \ref{fig:corr_simu}, case C-VD-VF. Some aspects still require further investigation, in particular why the NC-VD-VF configuration exhibits a monotonically increasing pattern, when Eq. \eqref{eq:scaling_IDelta_q} predicts no dependence on metaorder correlations. Nevertheless, we believe that the difference between C-NVD-VF (or NC-NVD-VF) and C-VD-VF supports and reinforces our claim that volume fluctuations coupled to volume dependence of the impact decay is key to account for such a non monotonic behaviour. 

\begin{figure}[H]
    \centering
    \includegraphics[width=\linewidth]{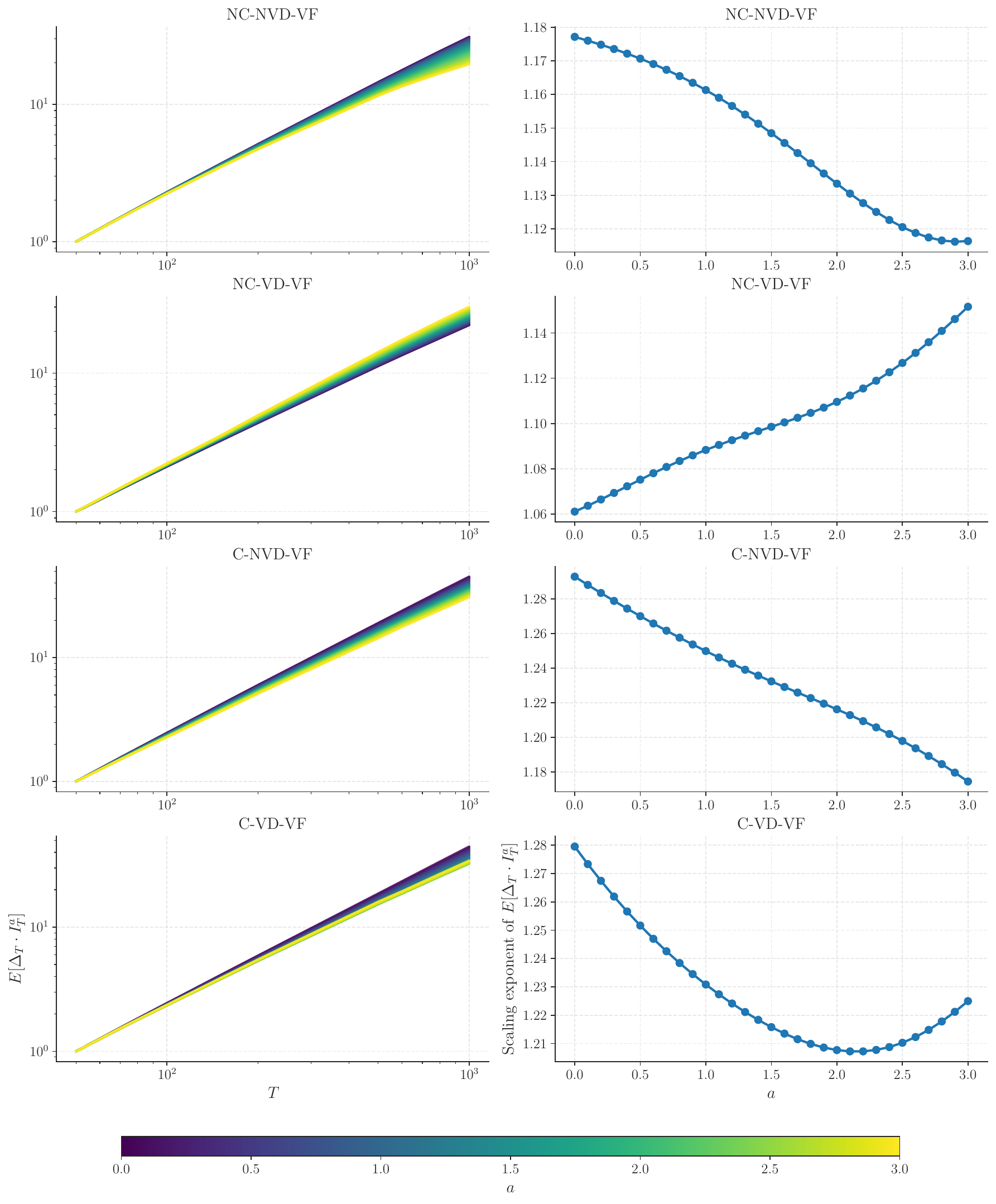}
    \caption{Covariance $(\Delta_T, I^a_T)$ as a function of $(T, a)$ for \textit{simulated} markets, with $m=3,\sigma_\ell=1, \lambda \sigma_\ell^2 = 1/8$ and $\lambda' =\lambda $, for the four configurations considered here. From top to bottom NC-NVD-VF, NC-VD-VF, C-NVD-VF and C-VD-VF. \textbf{Left:} Log-log plot of $\EX[\Delta_T \cdot I_T^a]$ vs. $T$ for different values of $a$. \textbf{Right:} Scaling exponents as a function of $a$, obtained by fitting the initial regime ($T < 10^3$).}
    \label{fig:cov_simu}
\end{figure}

\subsection{The correlation coefficient}

\begin{figure}[htbp]
    \centering
    \includegraphics[width=\linewidth]{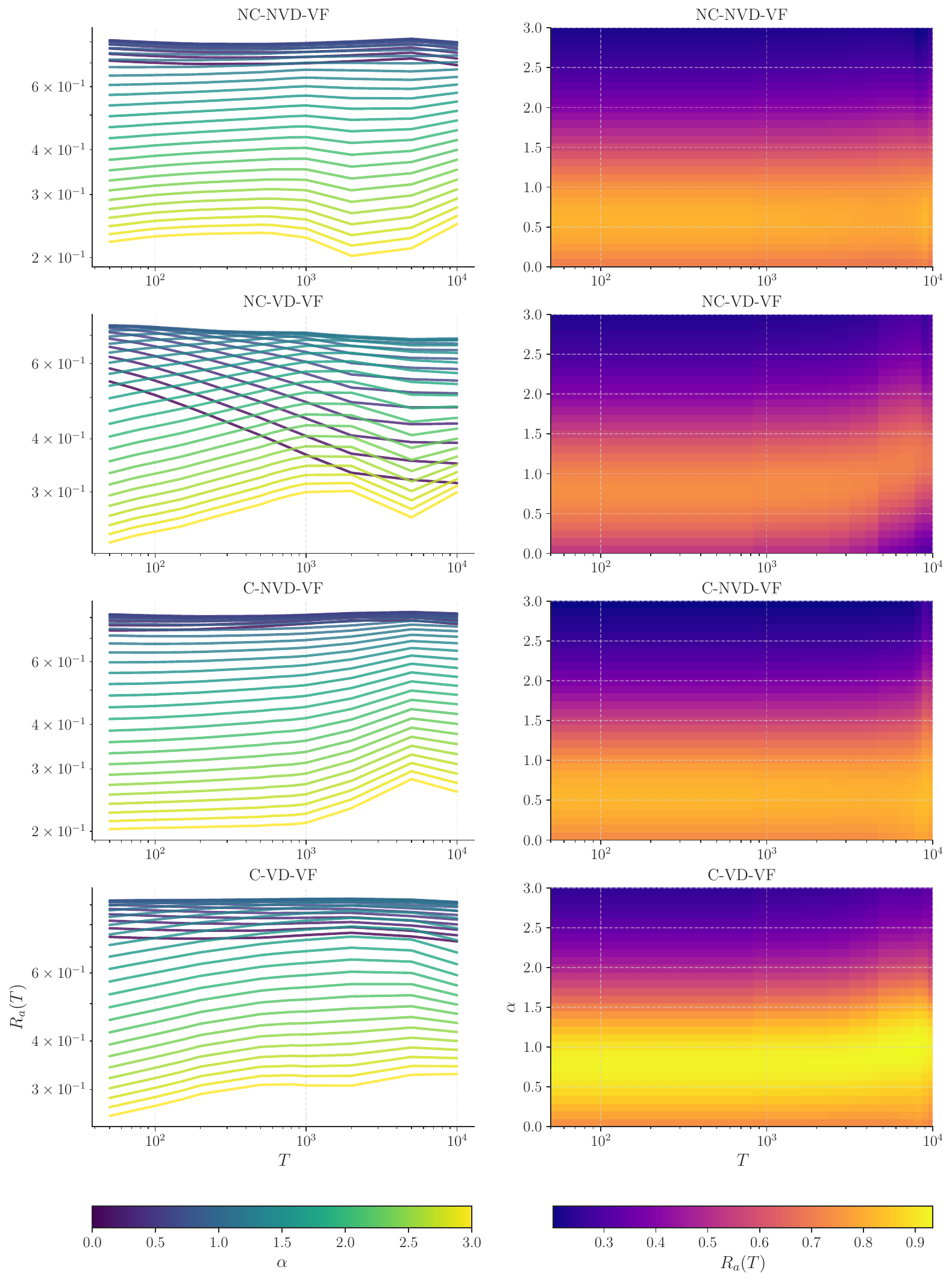}
    \caption{Simulations were done for $m=3,\sigma_\ell = 1$ and $\lambda = 1/(8\sigma_\ell)$. \textbf{Left column:} Evolution of the correlation for different values of $a$, showing the non monotonic behavior. \textbf{Right column:} Heatmap illustrating the distribution of correlation values within the $(a,T)$ space, indicating that the correlation reaches its peaks for $a \approx 0.5 - 1$, regardless of the $T$ values.}
    \label{fig:corr_simu}
\end{figure}

\begin{figure}[H]
    \centering
    \includegraphics[width=\linewidth]{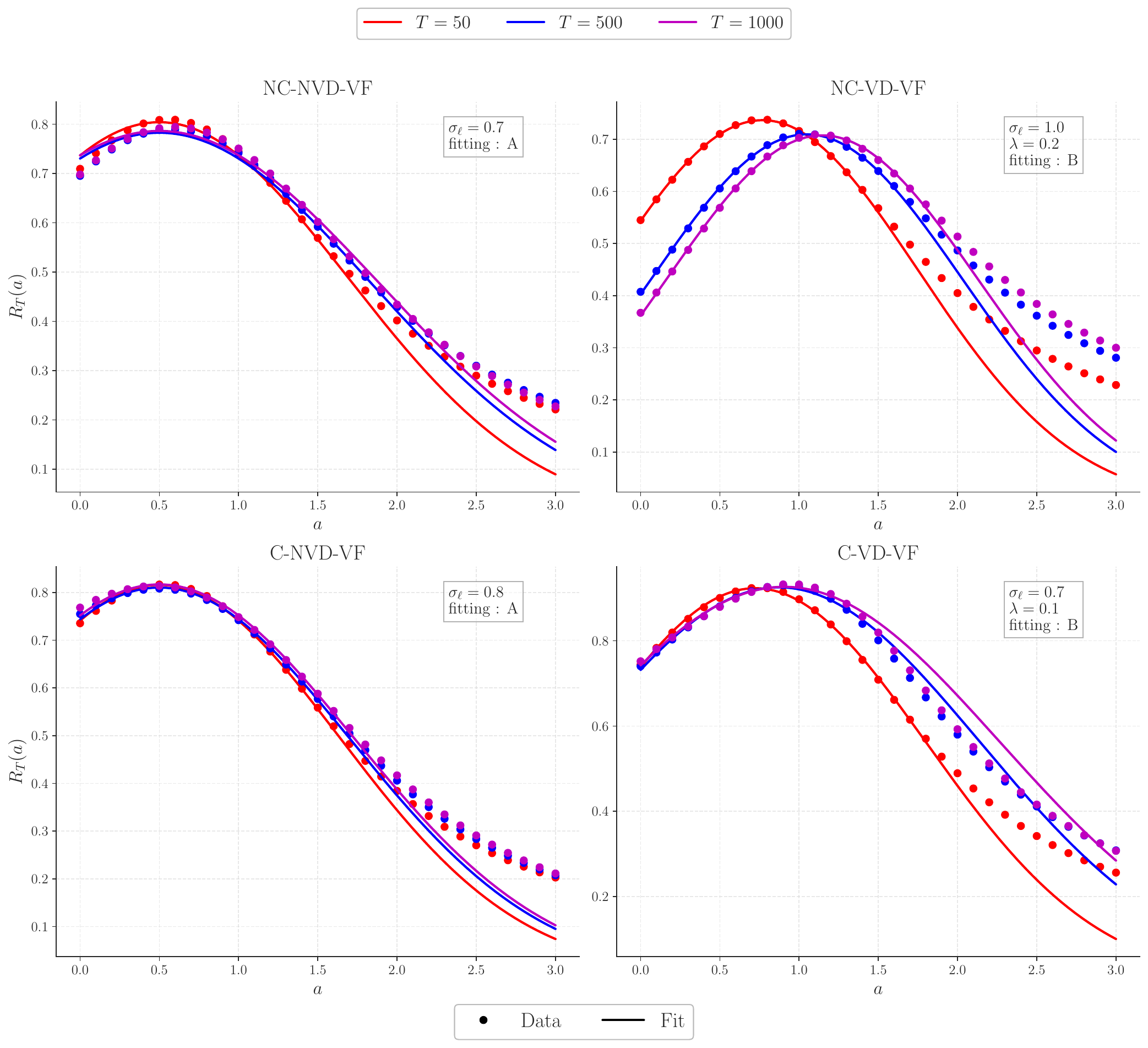}
    \caption{Fit of the correlation function $R_a(T)$ for several values of $T$. Since Eq.~\eqref{eq:R_a_vs_a} is valid only for $a < a_c$, the fit is restricted to $a < 1.5$. The letters (A, B) indicate which term of Eq.~\eqref{eq:R_a_vs_a} is being fitted. The empirical estimates of $\sigma_\ell$ and $\lambda$ obtained through this procedure are remarkably close to the values used as input in the simulations: $\sigma_\ell^2 = 1$, $\lambda = 1/8$.}
    \label{fig:R_aT_fitted}
\end{figure}

Finally, an important quantity is the correlation coefficient $R_a(T)$, for which our theoretical model also predicts a non-trivial behavior for fixed $T$ as a function of $a$. Once again, the resulting curves show quite a remarkable agreement with empirical data, as illustrated in Fig.~\ref{fig:corr_simu}. Moreover, by fitting Eq.~\eqref{eq:R_a_vs_a} to the simulated data, we can extract the values of $\sigma_\ell$ and $\lambda$, which are very close to the parameters originally used in the simulation, see Fig.~\ref{fig:R_aT_fitted}.

By fitting $R_a(T)$ as a function of $a$ for specific values of $T$, one can assess which term — $A$ or $B$ — is dominant, see Eq. \eqref{eq:R_a_vs_a}. This is done by successively fitting only one term at a time, i.e., either setting $B = 0$ and fitting $A$, or setting $A = 0$ and fitting $B$. Our theoretical framework also predicts which term should dominate depending on whether $\lambda \neq 0$.

Not only does the model show good qualitative agreement with the data, but the fits presented in Fig.~\ref{fig:R_aT_fitted} are also remarkably convincing from a quantitative point of view. In particular, we observe a clear match between:
\begin{itemize}
    \item the NVD case and fits using only the $A$-term (with negligible $B$),
    \item the VD case and fits where $B$ dominates (with $A$ negligible).
\end{itemize}
Moreover, the fits yield realistic estimates for both the input values of $\sigma_\ell$ and $\lambda$, further validating the consistency of the model.

\section{The puzzling effectiveness of proxy metaorders}

In this final section, we address a central puzzle in the study of price impact: the surprising effectiveness of ``proxy metaorders'' introduced in \cite{maitrier2025generatingrealisticmetaorderspublic}. Our algorithm constructs synthetic metaorders \textit{while preserving the exact trade history and sampling trades without replacement}, two conditions that turn out to be essential. This algorithm originates from a study of metaorder impact using the TSE dataset, which includes real trading identifiers. A striking initial finding was that randomly shuffling trading IDs and reconstructing synthetic metaorders still preserves the square-root impact law (SQL): see \cite{maitrier2025double} section 4.2 for details. 

However, one might argue that obtaining such a result relies on the prior knowledge of the original trading IDs. The shuffling process may preserve hidden information—such as the distribution of trading frequencies — which could, in turn, explain the impact function observed for the synthetic metaorders. Although appealing and somehow intuitive, this hypothesis was refuted in \cite{maitrier2025generatingrealisticmetaorderspublic} through the construction of synthetic metaorders using public data. Yet the justification of the success of that method in reproducing the SQL remained somewhat mysterious. 

The framework we introduce here allows one to justify further our proposal using purely simulated data. Although we have not yet been able to compute exactly the impact of proxy metaorders within our model, we believe that our numerical results are convincing enough to believe that the procedure proposed in Ref. \cite{maitrier2025generatingrealisticmetaorderspublic} is warranted. 

In Section \ref{sec:how_to_simulate}, we introduced a detailed procedure for generating a dataset that closely approximates the ideal case (such as the TSE dataset), which provides trade-by-trade data along with metaorder identifiers across the entire market. Building on this, we conduct a numerical experiment where we pretend we do not know the mapping between trades and metaorders, and construct a proxy in the spirit of \cite{maitrier2025generatingrealisticmetaorderspublic}. For the purpose of such an experiment, we assume no volume dependence, i.e, $\lambda = \lambda' = 0$\footnote{The proposed study and code can be readily extended to scenarios where $(\lambda,\lambda') \neq (0,0)$ and $q\sim LN(m,\sigma_q)$ by separating buy and sell orders, binning the volume $q$, and applying the algorithm using the corresponding value of $\mu_q$.
}. {Each metaorder can thus be characterized by only three parameters: its size $s$ drawn from a distribution $\Psi(s) \sim s^{-1-\mu}$, its execution rate which we choose to be the same for all metaorders $\tilde{\varphi} = \varphi$ and its average child order volume $q$, with $q\sim LN(m,\sigma_\ell^2)$}.

The core challenge in designing a reliable proxy for metaorders lies in aggregating market orders in a way that statistically approximates the true (yet usually unobservable) matching between traders and trades. A natural method to reconstruct realistic metaorders from the observed order flow is to first separate buy and sell orders and then, for each list, iterate through the orders while performing the following: if an order is already part of an existing metaorder, we skip it; otherwise, we draw a size $s \sim \psi(s)$ and group the next $s$ orders that occur within intervals of duration $\varphi$ into a new metaorder. Since splitting and grouping orders can bring together orders with the same sign that were actually executed far apart in time, we introduce an inter-time threshold between two child orders. If the inter-time is above the threshold, we consider that the two child orders belong to different metaorders. This inter-time constraint proves essential for reproducing the SQL, and corresponds to usual execution schemes where child orders tend to be relatively close to one another. A long pause in the execution schedule is tantamount to starting a new metaorder. 

\begin{figure}[H]
    \centering
    \includegraphics[width=0.7\linewidth]{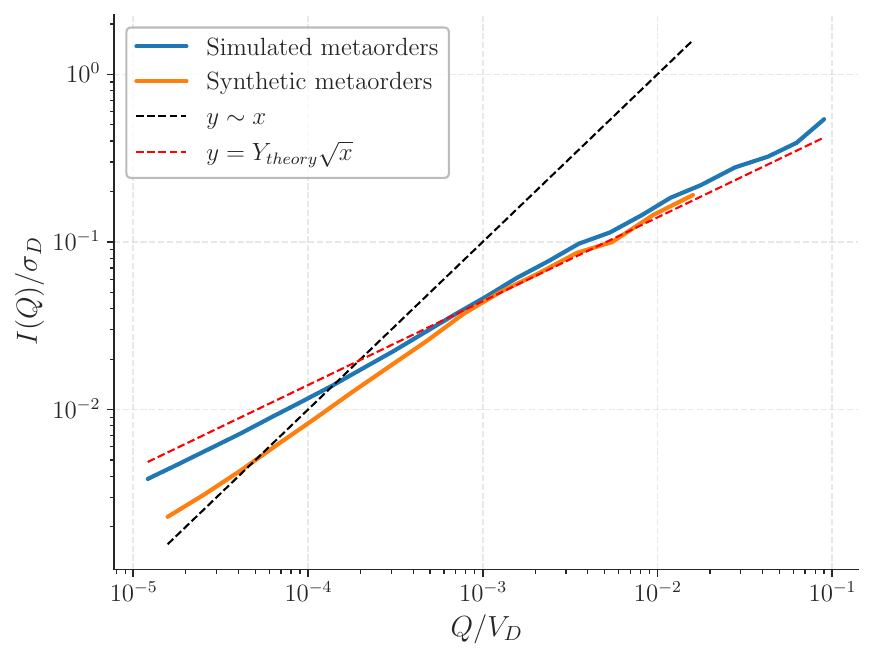}
    \caption{
Comparison of the impact of simulated metaorders in the C-NVD-VF setup and synthetic metaorders generated using the metaorder proxy and \textit{constructed from simulated prices}. For small $Q/V_D$, synthetic metaorders tend to have less concave, but after a crossover value around $10^{-3}$, it nicely converges to the expected SQL, which is an input of our simulation. {Note that we also recover the exact theoretical prefactor $Y_{\rm{theory}}$ computed in \cite{maitrier2025subtle}}. Both the simulation algorithm and the mapping function use $\varphi = 2.\, 10^{-3}$, $\mu = 1.5$, $m=3$ and $\sigma_\ell^2=1$. A total of $1{,}000{,}000$ simulated metaorders were generated. We obtain similar results in others simulations cases }

    \label{fig:sql_proxy}
\end{figure}

This procedure is summarized in Algorithm~\ref{algo:matching_meta}, where $C$ is a constant, which we arbitrarily set to $4\varphi$, as it provides satisfactory results, see Fig. \ref{fig:sql_proxy}, where we compare the numerical evaluation of the square-root law using the known exact matching between child orders and metaorders generated by our simulation (blue line) and the impact law estimated using proxy (or synthetic) metaorders (orange line). One sees that the agreement is almost perfect when $Q/V_D$ is not too small, whereas the effective behaviour of the
reconstructed impact becomes more linear. This is expected, since the start of short proxy metaorders have a higher probability to miss the start of ``real'' metaorders, for which the impact is most concave. {We also confirm that the derivation of the prefactor $Y$ of the SQL in \cite{maitrier2025subtle} is correct, namely $I(Q)=Y \sigma \sqrt{Q/V_D}$. We believe that this additional quantitative validation is important, since the prefactor is usually less studied in the literature, although it remains of significant interest for the estimate of actual impact costs.
}

These simulation results therefore bolster the claim made in \cite{maitrier2025generatingrealisticmetaorderspublic} that a realistic estimate of the impact of metaorders can be obtained using anonymous trade by trade data, provided the mapping function that generates proxy metaorders is chosen adequately.
In fact as shown in \cite{maitrier2025generatingrealisticmetaorderspublic} (Appendix), this mapping function, based on the theoretical framework developed here, also performs well on real data.

\begin{algorithm}[H]
    \caption{Generate Metaorder Identifiers with Time Threshold}
    \begin{algorithmic}[1]
        \Function{GenerateMetaIDs}{$t\_execs, \varphi, \mu, s_{\text{max}}$}
            \State $n \gets \text{len}(t\_execs)$
            \State $ids \gets \text{zeros}(n, \text{dtype=int})$
            \State $id\_meta \gets 1$
            \State $i \gets 0$
            \While{$i < n$}
                \State $size \gets \Psi(\mu, s_{\text{max}})$
                \State $count \gets 0$
                \State $current\_time \gets t\_execs[i]$
                \While{$count < size$ \textbf{and} $i < n$}
                    \State $ids[i] \gets id\_meta$
                    \State $count \gets count + 1$
                    \State $next\_time \gets current\_time + \text{Exp}(\varphi)$
                    \State $i\_next \gets \text{searchsorted}(t\_execs, next\_time, left)$
                    \If{$i\_next \geq n$ \textbf{or} $(t\_execs[i\_next] - next\_time) > C/\varphi$}
                        \State \textbf{break}
                    \EndIf
                    \State $i \gets i\_next$
                    \State $current\_time \gets t\_execs[i]$
                \EndWhile
                \State $id\_meta \gets id\_meta + 1$
                \While{$i < n$ \textbf{and} $ids[i] \neq 0$}
                    \State $i \gets i + 1$
                \EndWhile
            \EndWhile
            \State \Return $ids$
        \EndFunction
    \end{algorithmic}
    \label{algo:matching_meta}
\end{algorithm}

\section{Conclusion}

This work extends and complements our previous theoretical paper on the subtle interplay between impact, order flow and volatility \cite{maitrier2025subtle}. In that work, most of our predictions turned out to be in rather remarkable agreement with empirical observations, despite the simplifying mathematical approximations that we had to make. In the present paper, we show using numerical simulations that these approximations are actually quantitatively justified, which provides further support for the validity of our theoretical framework, and bolsters our conclusion that price volatility can be fully explained by the superposition of correlated metaorders that all impact prices, on average, as a square-root of executed volume. One of the most striking predictions of our model is the structure of the correlation between generalized order flow and returns, which is observed empirically and reproduced using our synthetic market generator.  

Furthermore, we were able to construct proxy metaorders from simulated order flow that reproduce the square-root law of market impact — a law that has long been, and in some circles still is, attributed to information revelation; see e.g. \cite{gabaix2006, hasbrouck2007empirical, saddier2024bayesian}. Our model, on the other hand, makes the assumption that impact is purely mechanical and a result of the random dynamics of latent liquidity that creates a buffer for price moves, see \cite{donier2015fullyconsistentminimalmodel, donier2016walras, bouchaud2018trades}.
The possibility of measuring the impact of metaorders from tape data (i.e. anonymized trades) was long thought to be impossible. However, Ref. \cite{maitrier2025generatingrealisticmetaorderspublic} showed that a suitable mapping between market orders and proxy metaorders allows one to reconstruct many statistical features of real metaorders. We confirm that this is indeed the case within our purely synthetic market as well, lending further credence to our  proposal \cite{maitrier2025generatingrealisticmetaorderspublic}.

The present framework not only confirms the validity of the theoretical analysis performed in \cite{maitrier2025subtle}, but also provides a useful market simulator that allows one to address a large number of interesting questions at the ``meso'' scale (as we do not zoom into the orderbook, fully ``micro'' dynamics) by simulating realistic trading environments. We hope that our codes, which are fully available here\footnote{\url{https://github.com/glatouille/ArtificialMarketSimulator.git}}, will be used and improved by both academic and professionals. 
Of course, several open questions remain. We propose here several directions for future research:
\begin{itemize}
    \item Calibrating the parameters of the model to reproduce quantitatively the short-term dynamics of the market, i.e., the full structure of the signature plot, Fig. \ref{fig:diffusion_two_timescale} and not only the long term diffusive behaviour. 
    \item Extending the simulations and the generalized propagator to include all order types, such as limit and cancellation orders, in order to model the full order book.
    \item Introducing heterogeneous trader categories (e.g., high- and low-frequency traders, market makers) and designing category-specific metaorders. For instance, market makers are likely to submit faster and smaller metaorders than low-frequency traders. This approach could shed light on the impact of different market participants.
    \item While the generalized propagator provides a convenient mathematical framework, some phenomenological aspects remain unexplained by the Latent Liquidity picture of Ref. \cite{donier2015fullyconsistentminimalmodel}. More work is needed to understand the mechanisms driving these effects in real markets (see also the discussion in \cite{maitrier2025double}).

    \item Investigating post-execution impact decay using proxy metaorders within our numerical model. It was suggested in Ref. \cite{maitrier2025generatingrealisticmetaorderspublic} that proxy metaorders built on market data can also reproduce the decay of real metaorders. Our framework is particularly well-suited to conduct an in-depth analysis of this phenomenon, in particular the role of the volume and duration of metaorders. 
    \item Finally, a precise analytical calculation of the impact of synthetic metaorders within our model would be extremely useful to distinguish the regime where such a procedure matches the square-root law of real metaorders from the small volume regime where a crossover towards a linear impact law would be obtained. 
\end{itemize}

\section*{Acknowledgments}

We wish to thank J. D. Farmer, J. Bonart, K. Kanazawa, F. Lillo, F. Patzelt, J. Ridgeway, M. Rosenbaum, A. Bugaenko, J. Kurth \& B. T\'oth for many enlightening conversations on these topics. This research was conducted within the Econophysics \& Complex Systems Research Chair, under the aegis of the Fondation du Risque, the Fondation de l'\'Ecole Polytechnique and Capital Fund Management.

\section*{Disclosure of interest}
The authors declare no conflicts of interest.

\section*{Funding}
No funding was received.

\printbibliography
\end{document}